\documentclass[runningheads,a4paper]{llncs}

\usepackage[T1]{fontenc}
\usepackage{amsmath,amssymb}
\usepackage{booktabs}
\usepackage{array}
\usepackage{graphicx}
\graphicspath{{figures/}{diagrams/}}
\PassOptionsToPackage{hyphens}{url}
\usepackage{url}
\usepackage[hidelinks]{hyperref}
\hypersetup{pdftitle={Basis-Only Funding Is Not Dividend Fair}}

\newcommand{\E}{\mathbb{E}}
\newcommand{\Prob}{\mathbb{P}}
\newcommand{\Var}{\operatorname{Var}}
\newcommand{\esssup}{\operatorname*{ess\,sup}}
\spnewtheorem{assumption}{Assumption}{\bfseries}{\itshape}

\title{Basis-Only Funding Is Not Dividend Fair:\\
Settlement Design in Crypto Equity Perpetuals}
\titlerunning{Basis-Only Funding Is Not Dividend Fair}
\author{Donghwa Seo\inst{1}\orcidID{0009-0003-5718-1535} \and
Doohwi Cha\inst{2}\orcidID{0009-0000-2513-6965} \and
Seunghan Son\inst{3}\orcidID{0009-0000-2902-9999} \and
Juyeong Lee\inst{4}\orcidID{0000-0002-4619-925X} \and
Minsuk Sung\inst{5}\orcidID{0009-0007-2282-6873} \and
Minjae Lee\inst{3}\orcidID{0000-0002-8937-9605}}
\authorrunning{D. Seo et al.}
\institute{DS Investment \& Securities, Seoul, South Korea\\
\email{properitas95@kaist.ac.kr}
\and Mirae Asset Securities, Seoul, South Korea\\
\email{doohwicha@gmail.com}
\and Independent Researcher, Seoul, South Korea\\
\email{qzpm5n@g.skku.edu, abraxasnz13@gmail.com}
\and EY Consulting, Seoul, South Korea\\
\email{juyeong.lee0839@gmail.com}
\and Korea University, Department of Artificial Intelligence, Seoul, South Korea\\
\email{minsuksung@korea.ac.kr}}

\begin{document}
\raggedbottom
\maketitle

\begin{abstract}
Crypto venues increasingly list perpetual futures on equities and other
real-world assets (RWAs), which trade continuously yet reference
dividend-paying stocks: a long equity perpetual marked to the ex-dividend cash
price bears the ex-dividend drop without receiving the dividend. We ask whether
the premium-index funding architecture these venues run --- funding set from
the contemporaneous basis $e_t=P_t-S_t$ rather than a carry target --- can
replicate shareholder cash flows for traders entering and exiting at arbitrary
times. It cannot. Funding serves two roles that coincide for non-dividend
crypto assets but split apart for equities: basis control and settlement. We
give the fair-funding identity, cast timing fairness as an optimal-stopping
condition, and characterize when the basis is a sufficient statistic for the
fair target. The decisive obstruction is measure-theoretic: a predictable
dividend is an atom, and no absolutely continuous funding path can replicate an
atom over holding intervals even in the weak limit, so fairness must match the
atom with an atom --- an event credit. Exchange-native data confirm this: at
each covered ex-date one venue posts a discrete funding atom equal to
$-D/(\mathrm{Mark}-D)$ while a standard premium-index venue posts none,
transferring the dividend across venues --- from native funding alone, with no
external spot.
\keywords{Equity perpetuals \and funding rates \and dividend settlement \and market design}
\end{abstract}

\section{Introduction}

Perpetual futures synthesize leveraged exposure without a fixed maturity: a
periodic funding payment replaces maturity-driven convergence. In crypto
markets the canonical design problem is basis control --- if the perpetual
trades rich to its reference index, longs pay shorts; if cheap, shorts pay
longs --- and the same architecture now appears in equity and RWA
perpetuals referencing single stocks, equity indices, and tokenized stocks.

The move from BTC to AAPL changes the economics. A stock has an ex-dividend
price, a shareholder cash-flow process, and corporate-action events. A
physical shareholder who sees the ex-dividend price fall by $D$---a classic
empirical regularity~\cite{EltonGruber1970}---also receives $D$; a long equity
perpetual marked to the ex-dividend cash price
does not own the share, so without a contract-side transfer the long bears
the drop without the cash flow and the short earns a predictable gain. The
question is a settlement-rule question: can the venue's funding mechanism
replicate the cash-flow economics of the reference asset for traders
entering and exiting at arbitrary times?

The textbook remedy is long-standing. Shiller's~\cite{Shiller1992} cash-settled perpetual
carries an explicit cash-flow index, so dividends are settled by construction.
Deployed crypto equity perpetuals, however, do not run Shiller's design: they
mark to the ex-dividend cash price and set funding from the contemporaneous
basis $e_t=P_t-S_t$, not from a carry target $r-q$. We therefore study the
architecture venues actually use --- the premium-index rule --- and ask when it
can nonetheless settle fairly. The question separates two roles of funding that
coincide for non-dividend crypto assets and split apart for equities:
\emph{basis control} (keeping $P$ near $S$) and \emph{settlement} (replicating
shareholder cash flows for arbitrary holding intervals). The theory delivers
four restrictions under which a basis-only rule can be fair, and each is
observable.

The question is blockchain market design, not generic mathematical
finance, because a perpetual's settlement rule is a programmable contract
object: on deployer-defined markets such as Hyperliquid HIP-3 the funding
formula, oracle, and any event credits are chosen and enforced in code,
and centralized venues face the same menu. Our results are constraints on
what any such contract can implement: continuous basis-only funding ---
the premium-index default --- cannot deliver dividend-fair settlement; an
event credit that matches the atom with an atom can.
In deployment these rules are published, not estimated: Hyperliquid sets funding to its own
premium index plus a clamped interest term~\cite{HyperliquidHIP3}, OKX and Gate use the same
standard premium-index rule~\cite{OKXStockTerms2026,GateFunding2026}, and Binance overlays a
one-time special dividend funding at each ex-date~\cite{BinanceDividendAdjustment2026,BinanceTradFiPerps2026}---so
the architecture our restrictions constrain is the deployed default, and the empirical
question is purely whether it settles the dividend.

Our analysis turns each role into a testable restriction. Weak (timing)
fairness is a martingale property of the hedged-book residual, and the fair
funding identity it implies depends on the predictable basis drift $\mu^e_t$,
not only on the carry wedge $r-q$ (Proposition~\ref{prop:martingale}). At a
predictable ex-dividend time fairness forces the basis itself to jump by the
dividend (Proposition~\ref{prop:atom}); and because a dividend is an atom, no
absolutely continuous funding path can replicate it uniformly over holding
intervals even when it converges weakly to the atom
(Proposition~\ref{prop:smearing}), so the right contract object is an event
credit, not a funding spike. The martingale condition also has an
optimal-stopping reading that pins the \emph{direction} of the unfairness: an
announcing-sequence argument shows an unmatched predictable dividend leaves
strictly positive, harvestable timing value for one side of the market
(Corollary~\ref{cor:announce}).

Finally we ask when the contemporaneous basis is enough on its own: we characterize when it
is a \emph{sufficient statistic} for the fair target (Proposition~\ref{prop:suffstat}), and
otherwise the best basis-only rule is a projection whose residual variance is a
misspecification floor (Proposition~\ref{prop:projection}). Under mean-reverting basis
dynamics the fair rule can \emph{decrease} in a positive premium while premium-index funding
increases in it (Proposition~\ref{prop:sign}).

Empirically, we test the observable implications on exchange-native data
alone---Binance and Hyperliquid funding, mark, and index, with no external
spot (Section~\ref{sec:empirics}). Three findings align with the theory.
(i) The deployed rule is the venue's published contemporaneous-premium function
(Hyperliquid's clamped premium rule; Binance's standard premium-index funding),
which we state rather than estimate. (ii) A dividend is matched by an atom or it
is not: at every covered ex-date Binance posts a discrete short-to-long funding
atom equal to $-D/(\mathrm{Mark}-D)$ (its published special dividend
funding), while a standard premium-index venue posts none. (iii) Consequently a
long held across the ex-date receives the
dividend on the event-credit venue and loses it on the standard one --- a
cross-venue settlement transfer of the dividend ($6$--$22$ basis points across
our names), read from native funding records with no external spot and no
arbitrage claim.

\section{Related Literature}

The paper connects four literatures. \emph{Perpetual-futures pricing.}
Shiller~\cite{Shiller1992} introduced the cash-settled perpetual with an explicit
cash-flow index, in which dividends are settled by construction; modern crypto-perp
theory instead develops no-arbitrage pricing, random-maturity representations, and
funding design for \emph{non-dividend} assets
\cite{AckererHugonnierJermann2025,HeManelaRossVonWachter2024,AngerisChitraEvansLorig2022,KimPark2025,Zhang2026Funding},
where funding has one job---basis control---and a parallel empirical strand documents
funding-rate arbitrage, the cash-and-carry basis trade, and cross-venue price
deviations~\cite{DaiLiYang2025PerpetualArbitrage,WerapunKarodeSuabootArpornthipSangiamkul2025,MakarovSchoar2020}.
We sit in the gap: deployed equity perpetuals inherit the premium-index rule but reference
an asset with Shiller's cash flows, so the role Shiller folded into the index reappears as
an unsolved settlement problem.

\emph{Tokenized equity and oracles.} A growing literature documents the dual-clock problem
and the oracle questions raised when tokenized stocks and RWAs trade outside cash-equity
hours~\cite{CongTokenizedStocks2025}; closely related work reconciles on-chain and off-chain
market data~\cite{YangKlagesMundtGudgeon2023}, asking how to \emph{measure} the reference
price, whereas we ask the orthogonal \emph{settlement} question once it is given.

\emph{TradFi equity derivatives} already carry the dividend machinery a perpetual must
reproduce without a maturity: forwards internalize dividends in the cost of carry,
single-stock futures use corporate-action adjustments~\cite{CMECorporateActions2026}, and
total-return swaps carry an explicit dividend leg whose stream is itself a developed
asset-pricing object~\cite{vanBinsbergenBrandtKoijen2012}. \emph{Venue documentation} shows
the design is unsettled in practice: Coinbase reflects dividends through
funding/mark~\cite{CoinbaseStockPerps2026,CoinbaseStockRisk2026}; Binance and Bitget document
special dividend funding~\cite{BinanceDividendAdjustment2026,BitgetDividendFunding2026}; OKX
applies standard stock-perp funding unless specified~\cite{OKXStockTerms2026}; Gate is
generic~\cite{GateFunding2026}; and Hyperliquid HIP-3 makes settlement
deployer-specific~\cite{HyperliquidHIP3}. Some
venues match the atom with an atom and most do not---the distinction our restrictions read off.

\section{Model}

Fix a filtered probability space
$(\Omega,\mathcal F,(\mathcal F_t)_{t\in[0,T]},\Prob)$ satisfying the usual
conditions, with $\Prob$ the pricing measure and $T<\infty$. Let $S_t$ be the
ex-dividend cash-equity price and $P_t$ the perpetual mark, both c\`adl\`ag
special semimartingales. Let $A_t$ be the cumulative shareholder distribution
process: c\`adl\`ag, nondecreasing, with absolutely continuous part
$a_t\,dt$ and jumps $\Delta A_\tau=D_\tau>0$ at predictable ex-dividend times
$\tau$. Let $r_t$ be a bounded predictable financing rate. The funding rule
$\phi_t$ is predictable, and funding is paid absolutely continuously: the long
perp pays $\phi_tP_t\,dt$ to the short.

For a position entered at $\tau_1$ and closed at $\tau_2$, the residual of the financed
spot-long/perp-short book is
\begin{equation}
R_{\tau_1,\tau_2}
=
\int_{\tau_1}^{\tau_2}(dS_t+dA_t-r_tS_t\,dt)
-\int_{\tau_1}^{\tau_2} dP_t
+\int_{\tau_1}^{\tau_2} \phi_tP_t\,dt .
\label{eq:basic_residual}
\end{equation}
When $P_t=S_t$ and $dA_t=q_tS_t\,dt$, zero drift requires the familiar
$\phi_t=r_t-q_t$. At a discrete dividend time $\tau$ at which $S$ and $P$ drop
by the same ex-dividend amount, the jump residual is $D_\tau$, and an explicit
event credit $C_\tau=D_\tau$, a total-return mark, or an equivalent settlement
adjustment is needed to remove it. This is the scalar benchmark; the paper
concerns the architecture that venues actually run.

The standard crypto architecture computes funding from a premium or basis. Let
\[
e_t=P_t-S_t,
\qquad
\phi_t=f(t,e_t),
\]
for a jointly measurable rule $f$; venue practice typically restricts further
to time-invariant rules $f(e_t)$. Substituting $P_t=S_t+e_t$ into
\eqref{eq:basic_residual} gives
\begin{equation}
R_{\tau_1,\tau_2}
=
(A_{\tau_2}-A_{\tau_1})
-(e_{\tau_2}-e_{\tau_1})
+\int_{\tau_1}^{\tau_2}\{f(t,e_t)P_t-r_tS_t\}\,dt .
\label{eq:basis_residual}
\end{equation}
Define the cumulative residual $R_t:=R_{0,t}$ and
\begin{equation}
M_t
=
A_t-e_t+\int_0^t\{f(u,e_u)P_u-r_uS_u\}\,du,
\qquad
R_{\tau_1,\tau_2}=M_{\tau_2}-M_{\tau_1} .
\label{eq:Mdef}
\end{equation}

\begin{assumption}\label{ass:reg}
(i) $A$, $e$, and the integrands in \eqref{eq:Mdef} are such that $M$ is a
special semimartingale of class~(D) on $[0,T]$, with $\E[A_T]<\infty$ and
every dividend atom $D_\tau$ integrable. (ii) The admissible stopping class is
$\mathcal T=\{\text{all stopping times with values in }[0,T]\}$, and
$\mathcal T_t=\{\rho\in\mathcal T:\rho\ge t\}$. (iii) Where stated, $A$ has
predictable density $a_t$ on its continuous part and $e$ has predictable
drift $\mu^e_t$ away from predictable jump times.
\end{assumption}

Class~(D) is the standard integrability for optional sampling over
$\mathcal T$~\cite{ProtterSDE2005} (it holds, e.g., when $S$, $P$, $A$ have integrable suprema
and $r$, $f$ are bounded); we state it once because every fairness
equivalence below uses it.

\begin{definition}[Weak and strong fairness]\label{def:fairness}
A funding rule is \emph{weakly fair} (timing-fair) if
$\E[R_\rho-R_\sigma\mid\mathcal F_\sigma]=0$ a.s.\ for all
$\sigma\le\rho$ in $\mathcal T$: no admissible entry/exit pair earns a
predictable carry. It is \emph{strongly fair} if $R_{\tau_1,\tau_2}=0$ pathwise for
every holding interval.
\end{definition}

\section{Theoretical Results}

Funding has two jobs that are distinct for equities: keep $P$ near $S$
(\emph{control}) and reproduce shareholder cash flows over any holding interval
(\emph{settlement}). We organize the analysis around three questions: what fair
funding requires in continuous time (Section~\ref{ssec:identity}); what happens at
the dividend itself, where the fair target is not a drift but an atom
(Section~\ref{ssec:atom}); and whether the deployed rule---a function of the
contemporaneous basis alone---can even match that target and keep its sign right
(Section~\ref{ssec:basis}). The martingale and optimal-stopping characterizations
are standard optional sampling, so we keep them brief; the load-bearing results
are the dividend obstruction and the basis-adequacy analysis.

\subsection{Fair funding is a drift, not a scalar}\label{ssec:identity}

By \eqref{eq:Mdef} the residual is $R_{\tau_1,\tau_2}=M_{\tau_2}-M_{\tau_1}$,
so weak fairness (Definition~\ref{def:fairness}) is exactly the optional-sampling
statement that $M$ is a martingale. Reading off the predictable part of its
decomposition turns that statement into a formula for funding.

\begin{proposition}[Fair funding identity]\label{prop:martingale}
Under Assumption~\ref{ass:reg}, a funding rule is weakly fair if and only if
$M$ is a (uniformly integrable) martingale. Away from predictable jump times the
fair rule is then
\begin{equation}
f(t,e_t)P_t=r_tS_t+\mu_t^e-a_t,
\qquad
f_t^{\mathrm{fair}}=\frac{r_tS_t+\mu_t^e-a_t}{P_t}.
\label{eq:fair_funding}
\end{equation}
\end{proposition}
\begin{proof}
Optional sampling makes ``$M$ a martingale'' equivalent to weak fairness via
\eqref{eq:Mdef}; the predictable finite-variation part of $M$ then vanishes,
and setting its density to zero gives \eqref{eq:fair_funding}.
\end{proof}

The equivalence is standard; the content is the identity, and three readings of
it organize the rest of the paper. First, the fair rule is \emph{not} the
textbook carry $r-q$: it is carry plus the predictable basis drift $\mu^e_t$
(and, through $a_t/P_t$, the dividend yield), so a rule calibrated to $r-q$ is
fair only in the knife-edge case $\mu^e_t\equiv0$. Second, the two jobs of
funding now visibly separate---$\mu^e_t$ describes where the \emph{basis} is
headed (control) while $r_tS_t-a_t$ is the cash-flow term (settlement)---and a
rule built to stabilize the basis and one built to settle cash flows coincide
only when $\mu^e$ is itself zero, the theme of Section~\ref{sec:sign}. Third,
$f^{\mathrm{fair}}$ is a function of \emph{predictable state}
$(r_t,S_t,P_t,\mu^e_t,a_t)$, not of the contemporaneous basis alone; whether a
deployed rule $f(e_t)$ can match it is therefore not automatic, and is the
question we take up in Section~\ref{sec:suff}. Everything downstream compares
the deployed premium rule against \eqref{eq:fair_funding}.

\subsection{The dividend atom cannot be funded continuously}\label{ssec:atom}

A dividend is the one place where the fair target is not a drift but an atom,
and it is where the premium-index architecture fails \emph{structurally} rather
than by miscalibration. The setup is immediate.

\begin{proposition}[Predictable dividend atoms]\label{prop:atom}
Let $\tau\in(0,T]$ be a predictable ex-dividend time with
$\Delta A_\tau=D_\tau$. Under Assumption~\ref{ass:reg}, weak fairness requires
\begin{equation}
\E[\Delta e_\tau\mid\mathcal F_{\tau-}]=D_\tau ,
\label{eq:jump_condition}
\end{equation}
and strong fairness requires $\Delta e_\tau=D_\tau$ pathwise. In particular, if
$P$ and $S$ drop by the same ex-dividend amount, so that $\Delta e_\tau=0$, no
absolutely continuous basis-only funding rule is fair across the event.
\end{proposition}
\begin{proof}
Funding is paid continuously, so $\Delta M_\tau=D_\tau-\Delta e_\tau$; optional
sampling along an announcing sequence $\sigma_n\uparrow\tau$ gives
$\E[\Delta M_\tau\mid\mathcal F_{\tau-}]=0$, which is \eqref{eq:jump_condition}.
\end{proof}

The statement is one line of algebra; its content is in the last sentence. A
venue that marks the perpetual to the ex-dividend cash price has $P$ and $S$
falling together at the ex-date, so $\Delta e_\tau=0$ while $D_\tau>0$ and the
condition \eqref{eq:jump_condition} is violated by the \emph{full} dividend.
The natural remedy---``spike funding just before the ex-date so accumulated funding
equals $D_\tau$''---fails for a reason central to the result, and the
reason is a choice of topology.

\begin{remark}[Event-window reading]\label{rem:window}
The same computation reads at the trade level: if the continuous integrands are
locally bounded near $\tau$, then for holding intervals
$[\tau_1^n,\tau_2^n]\ni\tau$ with $\tau_1^n,\tau_2^n\to\tau$, the residual
satisfies $R_{\tau_1^n,\tau_2^n}\to D_\tau-\Delta e_\tau$. An arbitrarily short
trade straddling the ex-date extracts the unmatched dividend---the unfairness is
not an artifact of long holding periods.
\end{remark}

\begin{proposition}[No absolutely continuous replication of the atom]
\label{prop:smearing}
Condition on $\mathcal F_{\tau-}$ so that $D_\tau=D>0$ is known, and let
$\nu=D\delta_\tau$ be the event-credit measure. For a finite signed measure
$\mu$ on $[0,T]$, define the interval discrepancy
$\Delta(\mu,\nu)=\sup_{a<b}\left|\mu((a,b])-\nu((a,b])\right|$. Then for every
$\mu$ absolutely continuous with respect to Lebesgue measure,
\begin{equation}
\Delta(\mu,\nu)\ge D,
\qquad\text{and}\qquad
\inf_{\mu\ll\mathrm{Leb}}\Delta(\mu,\nu)=D .
\label{eq:interval_gap}
\end{equation}
Writing $d\Lambda_t=dA_t-de_t+\{f(t,e_t)P_t-r_tS_t\}dt$, strong fairness on all
deterministic intervals is equivalent to $\Lambda\equiv0$ as a signed measure,
so an absolutely continuous funding term can cancel only the absolutely
continuous part of $dA_t-de_t$, never an unmatched singular atom.
\end{proposition}
\begin{proof}
For $I_\varepsilon=(\tau-\varepsilon,\tau]$, $\nu(I_\varepsilon)=D$ while
$|\mu|(I_\varepsilon)\to0$ by absolute continuity, so
$\Delta(\mu,\nu)\ge D-|\mu|(I_\varepsilon)\to D$; the choice
$\mu_\eta=D\cdot\mathrm{Unif}((\tau,\tau+\eta])$ attains
$\Delta(\mu_\eta,\nu)=D$. The final statement is uniqueness of the Lebesgue
decomposition.
\end{proof}

What matters is \emph{which} norm the result uses. Weak convergence---the sense in which a tall, narrow funding spike
``approximates'' a point mass---tests a measure against continuous functions,
and in that sense one can make funding converge to $D\delta_\tau$. But a
trader's P\&L over a holding interval $[\tau_1,\tau_2]$ integrates the funding
measure against the \emph{indicator} $\mathbf 1_{(\tau_1,\tau_2]}$, not against a
continuous test function. Equation~\eqref{eq:interval_gap} says that in the
topology trader P\&L actually inhabits---uniform discrepancy over
intervals---any absolutely continuous funding stream stays a full $D$ away from
the atom, however it is shaped or front-loaded: the gap is exactly $D$ for every
admissible smear, not a calibration error a finer spike would shrink. The design
consequence for venues is as follows. Dividend-fair settlement
cannot be delivered by \emph{any} continuous basis-only funding schedule; it
requires a mechanism carrying an atom of its own---an explicit event credit, a
dividend-adjusted or total-return mark, or an event-time funding payment of the
same singular type. Continuity of the funding rule and fairness across a
predictable dividend are mutually exclusive.

\begin{corollary}[No fair settlement atom off the price-adjustment instant]\label{cor:placement}
Let the dividend-induced price adjustment over $[0,T]$ be a finite measure $\pi$ with
$\pi([0,T])=D$---in the clean case $\pi=D\delta_\tau$ at the ex-drop, but in general
$\pi$ may carry several atoms (e.g.\ a declaration repricing and the ex-date drop) or be
spread over an interval (an overnight gap, or an oracle and mark adjusting at different
times). Let the settlement mechanism pay a long the measure $\mu$. The long's dividend
P\&L over a holding interval $(\tau_1,\tau_2]$ is $\mu((\tau_1,\tau_2])-\pi((\tau_1,\tau_2])$,
so fairness on \emph{every} holding interval holds iff $\mu=\pi$ as measures. In
particular a single settlement atom $\mu=D\delta_{\tau_s}$ is fair on every holding
interval iff $\pi=D\delta_{\tau_s}$, i.e.\ iff the price adjustment is itself a single
atom at exactly $\tau_s$. If $\pi$ is not a single atom then for any placement $\tau_s$
the interval discrepancy $\Delta(\mu,\pi)=\sup_{a<b}|\mu((a,b])-\pi((a,b])|$ equals the
mass of $\pi$ away from $\tau_s$, which is strictly positive: no perfectly fair atomic
instant exists.
\end{corollary}
\begin{proof}
Fairness on all intervals is $(\mu-\pi)((\tau_1,\tau_2])=0$ for all $\tau_1<\tau_2$,
equivalent to $\mu=\pi$ by additivity. For $\mu=D\delta_{\tau_s}$ this forces
$\pi=D\delta_{\tau_s}$; otherwise $\pi(\{\tau_s\}^{c})>0$ and an interval isolating that
mass gives $\Delta(\mu,\pi)\ge\pi(\{\tau_s\}^{c})>0$---the construction of
Proposition~\ref{prop:smearing} with $\nu$ replaced by $\pi$.
\end{proof}

The same martingale condition has an optimal-stopping reading that fixes the
\emph{direction} of the unfairness. Writing
$V_t^{\pm}=\esssup_{\rho\in\mathcal T_t}\E[\pm(R_\rho-R_t)\mid\mathcal F_t]$ for the
value of optimally timed exit on each side, optional sampling gives
$V^{+}=V^{-}=0$ iff $R$ is a martingale, so a one-sided failure names the side that
can profit~\cite{PeskirShiryaev2006,ProtterSDE2005}. This turns ``is the book
fair?'' into ``can either side harvest the dividend by timing the ex-date?'',
which the next corollary answers.

\begin{corollary}[Announcing-sequence barrier]\label{cor:announce}
Let $\tau$ be predictable with announcing sequence $\sigma_n\uparrow\tau$,
$\sigma_n<\tau$, and $\rho_n\downarrow\tau$ with $\rho_n>\tau$. Suppose the
continuous finite-variation terms of $R$ are dominated near $\tau$ by an
integrable random variable. If
$\E[D_\tau-\Delta e_\tau\mid\mathcal F_{\tau-}]\neq0$ on a set of positive
probability, then $\liminf_n V^+_{\sigma_n}>0$ or $\liminf_n V^-_{\sigma_n}>0$
there.
\end{corollary}
\begin{proof}
$R_{\rho_n}-R_{\sigma_n}=(D_\tau-\Delta e_\tau)+\xi_n$ with $\xi_n\to0$ in
$L^1$ by domination, so
$\E[R_{\rho_n}-R_{\sigma_n}\mid\mathcal F_{\sigma_n}]\to
\E[D_\tau-\Delta e_\tau\mid\mathcal F_{\tau-}]$; as
$\rho_n\in\mathcal T_{\sigma_n}$, $V^+_{\sigma_n}$ is bounded below by this limit
where positive (and the reverse direction where negative).
\end{proof}

Read with Proposition~\ref{prop:atom}, this says the unmatched dividend is not
an accounting residual that averages out: it is a strictly positive,
\emph{directional} timing premium that one side of the market can realize with
arbitrarily short trades around the ex-date. On a venue that marks to the
ex-dividend price ($\Delta e_\tau=0$) the harvestable side is the short---exactly
the party that gains the ex-dividend drop without paying the dividend. Whether a
venue closes this gap is what the native event evidence reads off directly
(Section~\ref{sec:empirics}): one venue matches the dividend atom, the
other does not.

\subsection{When the basis is not enough, and when it has the wrong sign}\label{ssec:basis}\label{sec:suff}\label{sec:sign}

Proposition~\ref{prop:martingale} makes fair funding a function of predictable
state, but venues run a function of the contemporaneous basis alone. The gap
between the two is what the data can test.

\begin{proposition}[Sufficient-statistic condition]\label{prop:suffstat}
Define the fair target $Y_t:=(r_tS_t+\mu_t^e-a_t)/P_t$ and suppose
\eqref{eq:jump_condition} holds at every predictable ex-dividend time.
(i) A weakly fair rule $\phi_t=f(t,e_t)$ exists iff $Y$ is, up to a
$(dt\otimes d\Prob)$-null set, measurable with respect to
$(t,\omega)\mapsto(t,e_t(\omega))$; equivalently, for $Y\in L^2$,
$\Var(Y_t\mid t,e_t)=0$ a.e. (ii) For the time-invariant rules $f(e_t)$ used in
venue practice the condition strengthens to $\Var(Y_t\mid e_t)=0$ under
$dt\otimes d\Prob$.
\end{proposition}
\begin{proof}
By Proposition~\ref{prop:martingale} weak fairness forces $f(t,e_t)=Y_t$ a.e.;
the Doob--Dynkin lemma on $([0,T]\times\Omega,\,dt\otimes d\Prob)$ gives the
stated measurability, and the conditional-variance form is its $L^2$ restatement.
\end{proof}

The proof is standard; the restatement is what makes the theory falsifiable and
is the bridge to the empirical section. ``Is a basis-only architecture
adequate?'' becomes ``does any predictable variable other than the basis carry
information about the fair target?''---and the observable residual
$g_t=\phi_t-(r_t-q_t)$ is a proxy whose conditional variance can be tested
directly. A non-zero $\Var(Y_t\mid e_t)$ means no contemporaneous-basis rule,
however shaped, can be fair; operationally, symbol identity, dividend yield,
ex-date proximity, and session state should \emph{not} predict $g_t$ once the
basis is controlled for, and if they do the architecture is rejected. Testing
this conditional-variance restriction natively requires a fitted $Y_t$, which we
leave to future work; the native evidence targets the sharper dividend
obstruction below.

\begin{corollary}[No universal premium rule]\label{cor:universal}
If two underlyings realize the same basis $x$ with positive
$dt\otimes d\Prob$ measure but different fair targets $Y_t$---different
financing, dividend density, expected basis drift, or corporate-action
exposure---then no single time-invariant $f(x)$, applied venue-wide, is weakly
fair for both.
\end{corollary}

This is a cross-sectional identification strategy: symbols whose
basis distributions overlap but whose dividend or drift profiles differ must
show different funding gaps at the same basis---something a single venue-wide
rule cannot produce. When exact fairness is out of reach, the same projection
geometry says how close one can get.

\begin{proposition}[Best basis-only rule is a projection]\label{prop:projection}
When exact fairness is infeasible, the quadratic-loss-optimal time-invariant
rule solves $\min_f\E\int_0^T P_t^2\{f(e_t)-Y_t\}^2dt$ and equals
\begin{equation}
f^\star(x)=\frac{\E[P_t^2\,Y_t\mid e_t=x]}{\E[P_t^2\mid e_t=x]},
\label{eq:projection}
\end{equation}
under the pooled measure $dt\otimes d\Prob$. The minimized loss is zero iff the
sufficient-statistic condition of Proposition~\ref{prop:suffstat}(ii) holds.
\end{proposition}
\begin{proof}
Time-invariant rules form the closed subspace of $L^2(P_t^2\,dt\,d\Prob)$
generated by measurable functions of $e_t$; orthogonal projection gives
$\E[P_t^2\{Y_t-f^\star(e_t)\}\mid e_t]=0$, which rearranges to
\eqref{eq:projection}.
\end{proof}

Beyond being the optimal rule, \eqref{eq:projection} is a \emph{design
benchmark}: its residual variance is the irreducible misspecification any
contemporaneous-basis-only venue must carry---a number a venue could estimate
and disclose. It also defines a concrete program: estimate $Y_t$, project it on
the basis, and read the leftover variance as the floor. Our native data motivate
this program but do not complete it---they identify the deployed premium rule and
the dividend atom directly (Section~\ref{sec:empirics}) but stop short of
estimating $Y_t$, the natural next step.

\medskip\noindent\textbf{Premium funding can have the wrong sign.}
The control/settlement split is sharpest when the basis mean-reverts, because
then stabilizing the basis and replicating cash flows point in opposite
directions.

\begin{proposition}[Sign reversal under mean reversion]\label{prop:sign}
Let $dA_t=qS_t\,dt$, write the relative basis $b_t=e_t/S_t$, and assume
\begin{gather*}
db_t=-\lambda b_t\,dt+\sigma_b\,dW^b_t,\qquad
dS_t=(r-q)S_t\,dt+\sigma_S S_t\,dW^S_t,\\
d\langle W^S,W^b\rangle_t=\rho\,dt .
\end{gather*}
Then the fair time-invariant rule satisfies
\begin{equation}
f^{\mathrm{fair}}(b_t)
=
(r-q)-\frac{\lambda b_t-\rho\,\sigma_S\sigma_b}{1+b_t},
\label{eq:sign_reversal}
\end{equation}
which for $\rho=0$ and small $b_t$ reads
$f^{\mathrm{fair}}(b_t)\approx(r-q)-\lambda b_t$. Whenever
$\lambda b_t>\rho\,\sigma_S\sigma_b$, fair funding at a positive basis lies
\emph{below} $r-q$ --- fair replication pays the short less than carry while
premium-index funding pays more --- and $f^{\mathrm{fair}}$ is
\emph{decreasing} in the premium whenever $\lambda+\rho\,\sigma_S\sigma_b>0$.
\end{proposition}
\begin{proof}
It\^o's product rule on $e_t=S_tb_t$ gives
$de_t=S_t\,db_t+b_t\,dS_t+d\langle S,b\rangle_t$, so the predictable drift is
$\mu^e_t=-\lambda b_tS_t+b_t(r-q)S_t+\rho\,\sigma_S\sigma_bS_t$. Substituting
$\mu^e_t$, $a_t=qS_t$, and $P_t=S_t(1+b_t)$ into \eqref{eq:fair_funding} and
simplifying yields \eqref{eq:sign_reversal}.
\end{proof}

This is the economic content of the control/settlement split. A positive premium
tells a control rule to make longs pay (to compress the basis); but if that
premium is expected to mean-revert, the short already collects the convergence,
so settlement fairness calls for paying the short \emph{less}, not more---fair
funding moves \emph{against} the premium while the premium-index rule moves with
it. Two cautions bound the claim. The level statement ($f^{\mathrm{fair}}<r-q$ at
a positive premium) and the slope statement (decreasing in the premium) are
distinct: differentiating \eqref{eq:sign_reversal} gives
$f^{\mathrm{fair}\prime}(b)=-(\lambda+\rho\,\sigma_S\sigma_b)/(1+b)^2$, so the
covariance term shifts both, and both hold only when mean reversion dominates.
And the conclusion is model-conditional on the OU form. Its precondition, however,
is now \emph{empirically} met: on venue-own premia (not the contaminated retail
basis) the premium mean-reverts for every Hyperliquid symbol---hourly
$\mathrm{AR}(1)$ of $0.72$ and half-life near two hours, so $\lambda>0$ is
estimated, not assumed (Section~\ref{sec:empirics})---while the deployed funding
slope on that premium is positive for all 13 names. These two model-free facts
(positive deployed slope, positive $\lambda$) are exactly the sign-reversal
precondition; only the resulting fair-slope sign is then read off the OU model. The
earlier appearance of a serially uncorrelated basis was an artifact of subtracting a
stale retail spot, not a property of the venue-own premium.

\section{Simulation: the settlement mechanism in a controlled market}\label{sec:sim}

The real event study below is low-powered---five Binance-covered ex-dates---and several
objects (a counterfactual fair-funding rule and the harvestable timing premium) are not
pinned down by any observational panel. We
therefore first exhibit the mechanism in a seeded Monte Carlo, calibrated to the published
Hyperliquid rule and to real venue-own premium statistics ($\sigma_b=6$ bp/hr, hourly
$\mathrm{AR}(1)=0.72$, half-life $2.1$ h, deployed loading $\kappa=0.05$, dividend $D=1\%$ of
price). Every quantity here is simulation, labelled as such; the real data in
Section~\ref{sec:empirics} then test the same mechanism.

\begin{figure}[t]\centering
\includegraphics[width=0.78\textwidth]{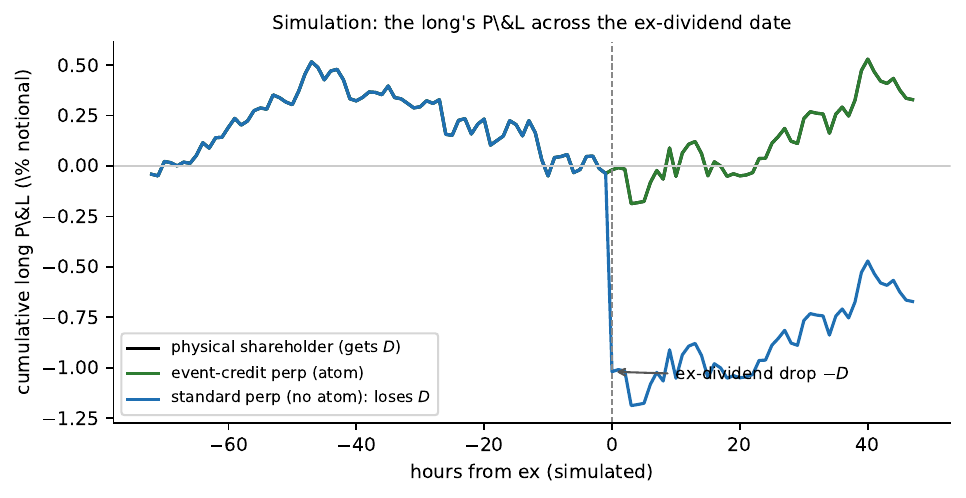}
\caption{Simulated long P\&L across the ex-dividend date.}
\label{fig:simprice}\end{figure}

\paragraph{The dividend drop and the long's P\&L (Fig.~\ref{fig:simprice}).} We give the
price its ordinary diffusion plus a mechanical drop of $D$ at the ex instant, mark the
perpetual to it, and accumulate the long's P\&L (price plus funding). A physical shareholder
is made whole: the $D$ drop is offset by the $D$ cash dividend. The standard premium-index
perpetual long, marked to the ex-dividend price with no compensating funding, ends a full
$D$ below---and the short a full $D$ above. The event-credit perpetual fires a $+D$ funding
atom to the long at the ex, which cancels the drop and restores the shareholder path. The
loss is not a transient that averages out: it is a permanent level shift of size $D$.

\begin{figure}[t]\centering
\includegraphics[width=0.78\textwidth]{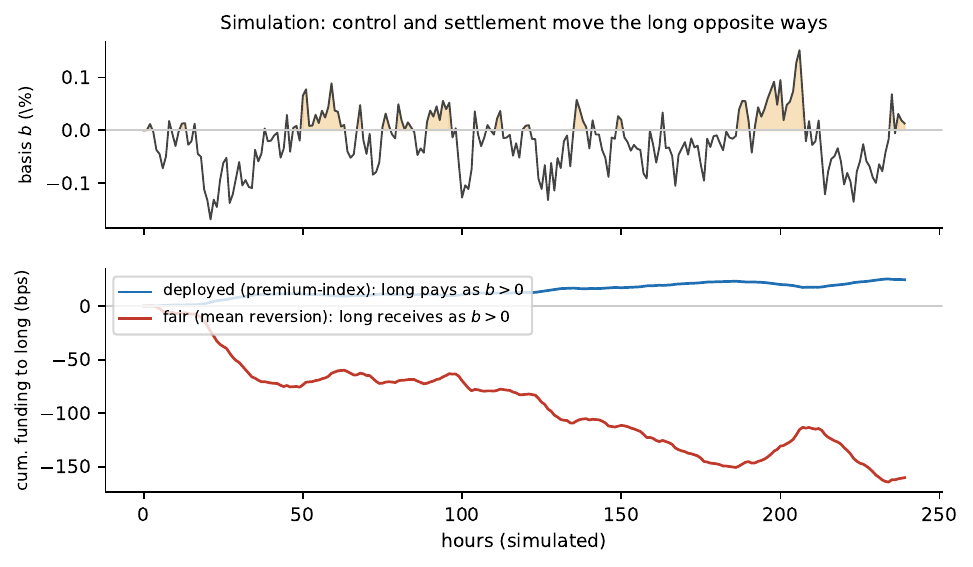}
\caption{Simulated control-versus-settlement sign reversal.}
\label{fig:simsign}\end{figure}

\paragraph{Control and settlement point opposite ways (Fig.~\ref{fig:simsign}).} On a
mean-reverting basis the deployed premium-index rule makes the long pay when the basis is
positive (pushing $P$ back toward $S$), whereas the fair rule pays the long in the same
state, because a positive basis predicts a downward correction. The two cumulative funding
curves move in opposite directions (Proposition~\ref{prop:sign}); the two inputs that drive
this---a positive deployed loading and $\lambda>0$---are exactly what the venue-own data
exhibit. A short straddle around the ex therefore harvests the unmatched dividend on the
standard venue (Corollary~\ref{cor:announce}).

\section{Empirical Evidence from Exchange-Native Data}\label{sec:empirics}

The theory predicts that a venue settles a predictable dividend fairly only if its funding
record carries a discrete atom of size $D$ at the ex
(Propositions~\ref{prop:atom}--\ref{prop:smearing}). We read this off deployed venues
directly, with no model fitting and no external price, by comparing a dividend-aware venue
(Binance, whose published rule fires a one-time special dividend funding) against standard
premium-index venues (Hyperliquid, OKX) that fire none. We use exchange-native data only:
Binance equity-perpetual funding and mark, the venue-own funding and premium of Hyperliquid
(13 symbols) and OKX, and the Bitget funding tape; each venue's premium is built from its own
mark and index, and every dividend test is anchored on the publicly announced dividend $D$
and ex-date---the only external input is the dividend calendar, never a price. Twelve
in-window ex-dates across seven names are covered; five carry both a Binance leg and a
standard-venue leg.

\paragraph{The atom is present on one venue, absent on the other.} At each covered ex-date
the Binance funding record contains a single entry equal to its published special
cash-dividend rule $-D/(\mathrm{Mark}-D)$, matching the announced dividend to
$|\Delta|<5\times10^{-6}$ (Table~\ref{tab:native}); Binance posts it at ex$-1$, 20:00 ET, the
evening before the ex-date and not at the announcement (the tape timestamp matches: NVDA's
atom at 2026-06-04 00:00\,UTC $=$ 2026-06-03 20:00 ET). The Hyperliquid and OKX standard
premium-index legs carry no such entry. The atom is identified not by magnitude---regular
equity funding is $O(3\text{--}12\text{ bp})$ per interval, of the same order---but by its
equality to $-D/(\mathrm{Mark}-D)$ at the ex-timestamp. Bitget settles dividends by a
snapshot credit off the funding tape, so no $D$-sized entry appears in its funding history:
the same settlement economics on a different accounting surface.

\begin{figure}[t]\centering
\includegraphics[width=\textwidth]{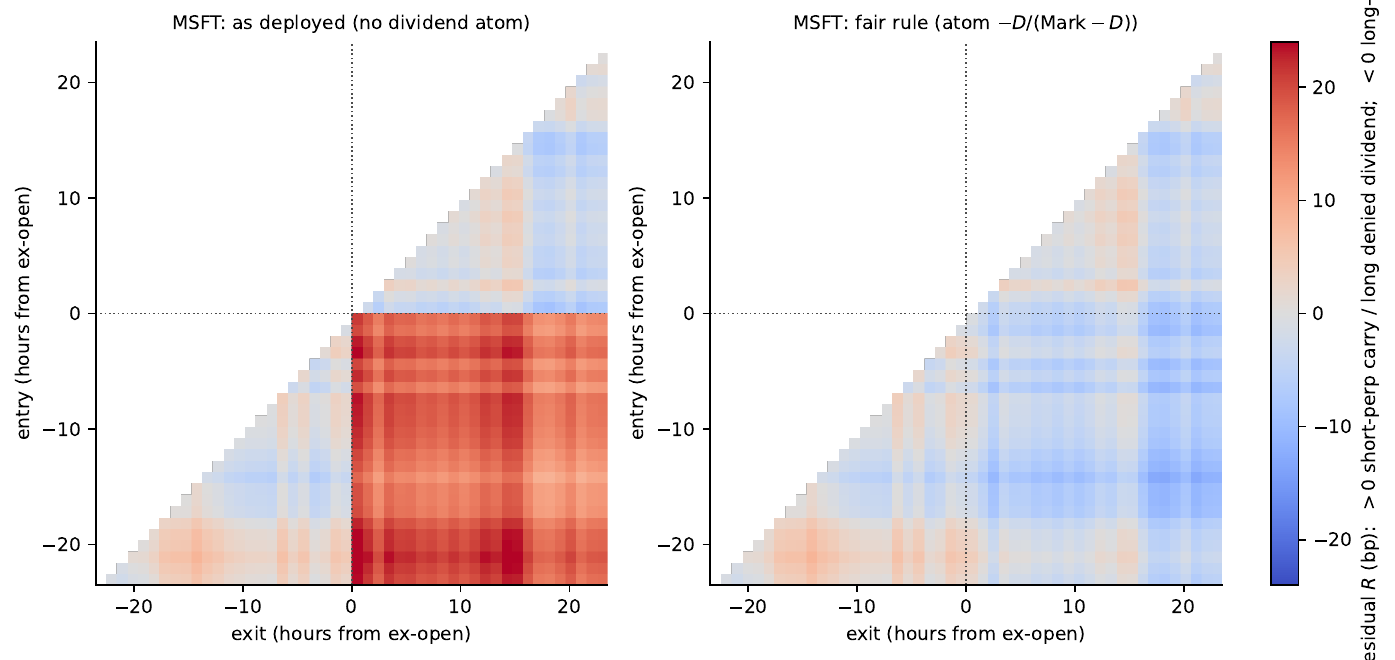}
\caption{Fairness residual $R$(entry, exit): as deployed vs.\ fair rule (MSFT).}
\label{fig:fairness}\end{figure}

\paragraph{What the missing atom does to entry/exit fairness.} The consequence for a holder
is cleanest in the residual of the financed spot-long/perp-short book, $R(t_1,t_2)$ of
\eqref{eq:basis_residual}, built from native funding, mark, and index for every entry $t_1$
and later exit $t_2$ in a $\pm24$-hour window around the ex (Fig.~\ref{fig:fairness}, MSFT).
As deployed (left), Hyperliquid posts no atom, so every entry-before-ex / exit-after-ex pair
inherits the whole dividend: $R$ is a uniform positive block---a mean of $+17$ bp over
straddling pairs, against $D/\mathrm{Mark}=21.7$ bp net of the ambient basis---a
\emph{predictable} carry to the short-perp side, which is exactly the dividend the long perp
is denied. Pairs that do not straddle the ex sit at the ambient baseline ($-0.5$ bp). Posting
the proposition's atom $-D/(\mathrm{Mark}-D)$ (right)---the published special-dividend
rule---collapses the block onto that baseline ($-4.5$ bp): the predictable dividend carry is
removed and weak (timing) fairness is restored. The faint residual speckle ($\pm3.8$ bp, mean
zero) is basis risk, not unfairness---strong pathwise fairness is unattainable for any
basis-funded perpetual, and any small tilt is the $\mu^e$ ambient carry of
\eqref{eq:fair_funding}, separate from the dividend. The block scales with $D/\mathrm{Mark}$
across names (ORCL $36$, MSFT $22$, NVDA $12$ bp) and vanishes for NVDA's token \$0.01
dividend ($0.5$ bp), a built-in no-event control. Its cross-venue counterpart is the published
atom itself (Table~\ref{tab:native}); this is a settlement-leg size, not an arbitrage net of
fees~\cite{DaiLiYang2025PerpetualArbitrage,WerapunKarodeSuabootArpornthipSangiamkul2025}.

\begin{figure}[t]\centering
\includegraphics[width=0.84\textwidth]{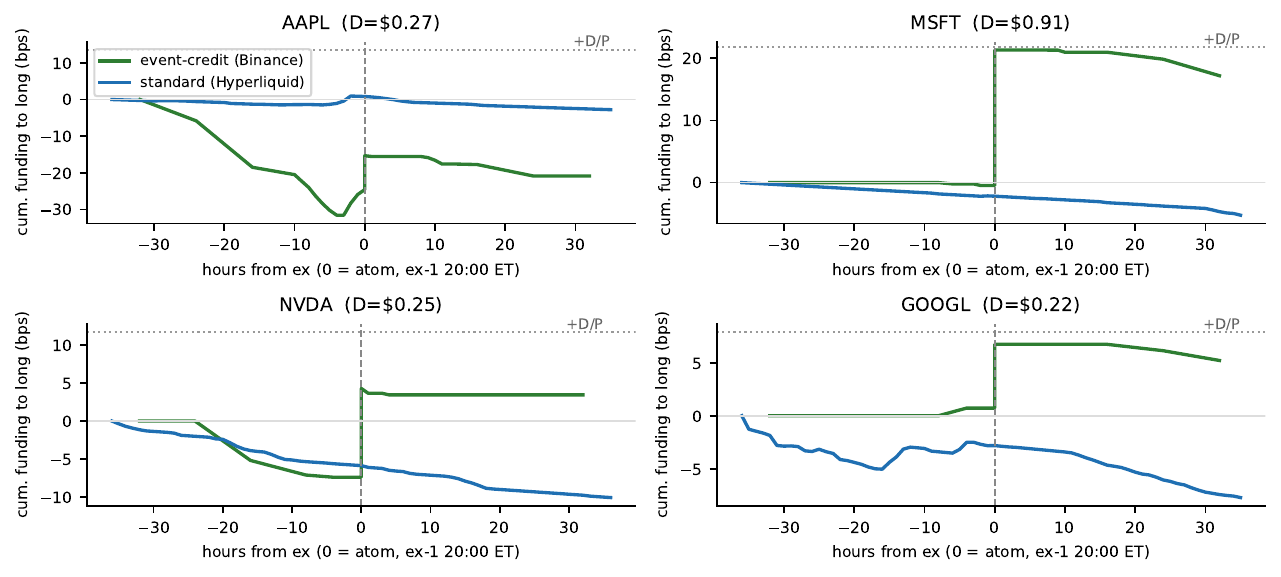}
\caption{Long-vs-short funding across ex-dates.}
\label{fig:favor}\end{figure}

\paragraph{Carry is a separate, symbol-dependent question.} The dividend loss is distinct from
ordinary carry. Across the thirteen Hyperliquid names mean funding is positive---longs pay on
average and the perpetuals trade rich---yet every name has long-favoured sub-windows
(Fig.~\ref{fig:favor}) where the perpetual sits at a discount (a cumulative long-favoured
window of $+294$ bp on TSM, $+92$ on
COST, $+70$ on AAPL). Whether a long's \emph{carry} pays therefore depends on symbol and
timing; whether its \emph{dividend} leg is fair does not---on a standard venue the cross-ex
long always loses $D/\mathrm{Mark}$, whatever the carry sign.

The design is deliberately narrow and the claims are qualitative. Five Binance ex-dates carry
the event tests (Binance equity coverage begins at listing, and META lacks a Hyperliquid leg
in the window); OKX and Bitget enter as additional standard and event-credit comparators. The
ex-open price drop is not separable from ambient volatility at hourly resolution---the dividend
is a few basis points against $\sim$$6$ bp hourly moves---so the evidence rests on the funding
atom, not a measured price step, and the residual heatmap is legible only where $D/\mathrm{Mark}$
clears the basis noise (clean for the larger dividends, basis-dominated for the smallest). With
the fair slope not estimated natively, the sign reversal stays model-conditional, and a
scheduled, synchronized venue panel over more ex-dates is the natural next step.

\begin{table}[t]\centering
\caption{Native ex-date results.}
\label{tab:native}
\begin{tabular}{llrrrr}
\toprule
Symbol & Ex-date & $D$ & $\phi_{\mathrm{atom}}$ & $-D/(\mathrm{Mark}-D)$ & $D/P$ (bp) \\
\midrule
AAPL  & 2026-05-11 & 0.27  & $-0.00092$ & $-0.00092$ & 9.2 \\
MSFT  & 2026-05-21 & 0.91  & $-0.00218$ & $-0.00217$ & 21.8 \\
NVDA  & 2026-06-04 & 0.25  & $-0.00117$ & $-0.00117$ & 11.7 \\
GOOGL & 2026-06-08 & 0.22  & $-0.00060$ & $-0.00060$ & 6.0 \\
META  & 2026-06-15 & 0.525 & $-0.00091$ & $-0.00091$ & 9.1 \\
\bottomrule
\end{tabular}
\end{table}

\section{Discussion}

\paragraph{Control versus settlement.} A positive premium makes longs pay
shorts (compressing the basis), but under mean reversion fair settlement can
require paying the short \emph{less} (Proposition~\ref{prop:sign}). The two
objectives coincide only in a non-generic case: a venue that trades settlement fairness
against basis control through an HJB criterion (a quadratic settlement loss plus a
basis-control penalty, under controlled basis dynamics) has optimal feedback
$f^\star=Y+\tfrac{\beta}{2P^2}\,\partial_eV$, equal to the settlement-fair $Y$ only where
the shadow value $\partial_eV$ of moving the basis vanishes; the native evidence
(Section~\ref{sec:empirics}) shows the deployed rule is a contemporaneous-premium
function, so it optimizes basis control, not settlement, and no single venue-wide rule
equalizes symbols with different fair targets (Corollary~\ref{cor:universal}).

\paragraph{What venues can do.} A venue that marks to the ex-dividend cash price and uses
only continuous basis funding cannot be dividend-fair for arbitrary holding intervals
(Propositions~\ref{prop:atom}--\ref{prop:smearing}); fairness is restored by any mechanism
that matches the atom with an atom---an explicit event credit, a total-return or
dividend-adjusted mark, an event-time funding payment, or a dividend accrual account. This
is observed natively: Binance's special dividend funding appears as a discrete
short-to-long atom equal to $-D/(\mathrm{Mark}-D)$ at each ex-date~\cite{BinanceDividendAdjustment2026}, the
atom-for-atom correction the propositions require, while a standard premium-index venue
supplies none and its long loses $D/P$; Bitget documents a comparable
snapshot credit~\cite{BitgetDividendFunding2026}, and OKX's standard stock-perp language is
the contrasting design~\cite{OKXStockTerms2026}. When exact fairness fails,
Proposition~\ref{prop:projection} gives the benchmark: project $Y_t$ on the basis and read
the residual variance as the irreducible misspecification of any basis-only rule.

\section{Conclusion}

A long equity perpetual marked to an ex-dividend price has a missing
settlement leg, and we characterized when funding can repair it: the residual
must be a martingale, predictable dividend atoms must be matched by basis jumps
or explicit event mechanisms, the basis must be a sufficient statistic for the
carry wedge, and absolutely continuous funding cannot replicate an atom over
holding intervals. On exchange-native data the deployed rule is a
contemporaneous-premium function, a dividend is matched by a discrete funding
atom on one venue and by nothing on another, and a long held across the ex-date
loses the dividend on the standard venue --- the settlement transfer the theory
predicts, read directly from native funding.
Funding rules in equity perpetuals are not only the price of leverage; they are settlement
mechanisms, and should be engineered---and tested---as such. Extended proofs, full panels,
and the data-provenance record and scripts are in the supplementary companion~\cite{RQ1Harness2026}.

\bibliographystyle{splncs04}
\bibliography{references}

\end{document}